\documentclass[aps,pra,twocolumn,superscriptaddress,showpacs]{revtex4}

\usepackage{graphicx}
\begin{document}

\title{Nonlocal information as condition for violations of Bell inequality and information causality}

\author{Yang Xiang}
\email{njuxy@sohu.com}
\affiliation{School of Physics and Electronics, Henan University, Kaifeng 475004, Henan province, China }

\author{Shi-Jie Xiong }

\affiliation{National Laboratory of Solid State Microstructures and
Department of Physics, Nanjing University, Nanjing 210093, China}
\date{\today}
\begin{abstract}
On the basis of local realism theory, nonlocal information is necessary for violation of Bell's inequality. From a theoretical point of view, 
nonlocal information is essentially the mutual information on distant outcome and measurement setting. In this work we prove that
if the measurement is free and unbiased, the mutual information about the distant outcome and setting is both necessary for the violation of Bell's inequality in the case with unbiased marginal probabilities. In the case with biased marginal probabilities, we point out
that the mutual information about distant outcome cease to be necessary for violation of Bell's inequality,
while the mutual information about distant measurement settings is still required.
We also prove that the mutual information about distant measurement settings must be contained in the transmitted messages due to the freedom of measurement choices. Finally we point out that the mutual information about both distant outcome and measurement settings are necessary for a violation of information causality.

\end{abstract}

\maketitle




The local realism theory (LRT) states that physical systems
can be described by local objective properties (physical reality)
which are independent of observation. Violation of Bell's inequalities \cite{bell,chsh} suggests that
the quantum mechanics cannot be regarded as LRT. We can also say that Bell's inequality sets a bound to correlations which
can be produced between separate classical subsystems if there is no communication between them. So a fundamental
information-theoretic question naturally arises: How much information must be communicated between two classical subsystems
in order to cause violation of Bell's inequality? We may call such information about distant partner's measurement settings and outcomes
as nonlocal information.
In \cite{toner}, Toner and Bacon proved that just one bit of message is sufficient to produce correlations
of measurements on an entangled Bell pair state. A one bit message is a strong nonlocal resource, it not only can
produce correlations of measurements on a Bell pair but also can simulate nonlocal box proposed by Popescu and Rohrlich (PR) \cite{pr} which is a
hypothetically stronger
nonlocal resource than a Bell pair \cite{cerf} but still preserves the non-signaling condition. Although a one bit message
is a much stronger nonlocal resource than a Bell pair, it must still be subtly designed, otherwise it cannot cause violation
of Bell's inequality. So the properties that nonlocal information must have in order to violate Bell's inequality attract much
investigation \cite{shimony}.

Recently Paw{\l}owski \emph{et al.} \cite{pawlowski1} provided a novel analysis of nonlocal information. They proved that in the condition of
maintaining realism and the experimenter's freedom of choosing measurement settings it is impossible
to achieve a violation of Bell's inequality if nonlocal information does not contain information about both the measurement setting and the
outcome of the distant parter. They formulated their conclusion in terms of a conception named ``guessed information'' $J(X\rightarrow Y)$,
which represents the average probability to correctly guess random variable $Y$ by knowing the value of $X$.
After expressing Bell's inequality in terms of these guessed information they found that if either one of $J(\lambda,\chi\rightarrow B)$ and
$J(\lambda,\chi\rightarrow b)$ equals $1/2$, violation of Bell's inequality would not occur, where $\lambda$ is the shared hidden variable,
$\chi$ is the message which Bob sends to Alice, and $B$ and $b$ respectively represent the outcome and measurement setting at Bob's site.

In this work we elaborate this issue by adopting the commonly used information theoretic conception, the mutual information.
Nonlocal information is contained in hidden variable $\lambda$ and message $\chi$, and in fact it is the mutual information
that relates them to the outcome and setting at the distant one. We will present the bound conditions of the probability that Alice can have a correct guess
in terms of mutual information by using the Fano inequality \cite{cover}. Furthermore, we will
study the question whether a violation of Bell's inequality occurs for any given amount of mutual information. Our method can provide a direct relation
of the violation of Bell's inequality to the nonlocal information (or, say, the mutual information). We also prove that if either one of the mutual information about
the distant outcome and
about the measurement setting equals zero, the violation of information causality \cite{pawlowski2}
would be impossible.

We first give a brief introduction of the Fano inequality \cite{cover}. Suppose we know the value of a random variable $Y$ and wish to guess another random variable
$X$. The Fano inequality relates the probability of guessing in error to the conditional entropy $H(X|Y)$:
\begin{eqnarray}
H(P_{e})+P_{e}\log{(|\chi|-1)}\geq H(X|Y),
\label{fanoineq}
\end{eqnarray}
where $P_{e}$ is the probability of guessing in error, $H(P_{e})=-P_{e}\log{P_{e}}-(1-P_{e})\log{(1-P_{e})}$,
and the conditional entropy can be written in terms of the mutual information between $X$ and $Y$: $H(X|Y)=H(X)-I(X;Y)$. In Eq. (\ref{fanoineq})
$|\chi|$ denotes the number of alphabet which $X$ can take. Here we just discuss the case of $|\chi|=2$, so Eq. (\ref{fanoineq}) can
reduce to
\begin{eqnarray}
H(P_{e})\geq H(X|Y).
\label{fano}
\end{eqnarray}

In the present case we express CHSH inequality in terms of the probabilities,
\begin{eqnarray}
\frac{1}{4}\sum^{1}_{a,b=0}{P(A\oplus B=ab|a,b)}\leq\frac{3}{4},
\label{bellineq}
\end{eqnarray}
where $\oplus$ denotes addition modulo $2$, and $P(A\oplus B=ab|a,b)$ is the conditional probability with which $A,B$ satisfy $A\oplus B=ab$ when
$a,b$ have been given. In the experiment of Bell's inequality we assume that both Alice and Bob can freely and unbiasedly choose their measurement
settings, $a,b=0,1$ and $A,B=0,1$, denoting their outcomes respectively. Since we assume that $P(a)=P(b)=1/2$ for any $a,b$, Eq. (\ref{bellineq})
can be transformed into the following form which is the CHSH inequality from Alice's perspective:
\begin{eqnarray}
&&\frac{\sum_{b}{P(A=B|a=0,b)}}{4}+\frac{\sum_{b}{P(A=B\oplus b|a=1,b)}}{4}\nonumber\\
&=&\frac{P(A=B|a=0)}{2}+\frac{P(A=B\oplus b|a=1)}{2}\leq\frac{3}{4},
\label{bellineq1}
\end{eqnarray}
where $P(A=B|a=0)$ is the conditional probability that Alice's outcome $A$ is equal to Bob's outcome $B$ when Alice has
chosen the measurement setting $a=0$, and it's similar for $P(A=B\oplus b|a=1)$.
So in order to achieve violation of Bell's inequality Eq.(\ref{bellineq1}) Alice's mission is maximizing
$P(A=B|a=0)$ and $P(A=B\oplus b|a=1)$ to the best of her abilities. The resource that Alice can utilize are
hidden variable $\lambda$ and message $\chi$ which Bob send to her. By using Eq.(\ref{fano}), we can
bound the two probabilities in terms of mutual information $I(B;\lambda,\chi)$ and $I(B\oplus b;\lambda,\chi)$:
\begin{eqnarray}
&&H[1-P(A=B|a=0)]\nonumber\\
&&\geq H(B|\lambda,\chi)=H(B)-I(B;\lambda,\chi);
\label{fano1}\\
&&H[1-P(A=B\oplus b|a=1)]\nonumber\\
&&\geq H(B\oplus b|\lambda,\chi)=H(B\oplus b)-I(B\oplus b;\lambda,\chi).
\label{fano2}
\end{eqnarray}
Here we assume that the marginal probabilities are unbiased, that means $H(B)=H(B\oplus b)=1$. We will come back
to the case with biased marginal probabilities later, in that case we find the mutual information about distant outcome cease to be necessary for violation of Bell inequality
while the mutual information about distant measurement settings is still required.
So Eq.(\ref{fano1}) and Eq.(\ref{fano2}) can reduce to(noticing $H[1-P(A=B|a=0)]=H[P(A=B|a=0)]$, etc) :
\begin{eqnarray}
&&H[P(A=B|a=0)]\geq 1-I(B;\lambda,\chi);
\label{fano3}\\
&&H[P(A=B\oplus b|a=1)]\geq 1-I(B\oplus b;\lambda,\chi).
\label{fano4}
\end{eqnarray}

Now we denote $\frac{P(A=B|a=0)}{2}+\frac{P(A=B\oplus b|a=1)}{2}$ as $\beta$, only when $\beta_{max}>3/4$ it is possible to violate Bell's inequality.
Considering the bound condition Eq.(\ref{fano3}) and Eq.(\ref{fano4}), it's obvious that when $I(B;\lambda,\chi)=0$
the maximum of $P(A=B|a=0)$ is $\frac{1}{2}$, therefore $\beta_{max}$ can not exceed $3/4$.
In this case it is impossible to achieve a violation of Bell inequality.

We are now in the position to prove that the information of Bob's
measurement settings also must be available at Alice site ($I(b;\lambda,\chi)>0$) in order to
achieve $\beta_{max}>3/4$. We will prove if $I(b;\lambda,\chi)=0$ then $\beta_{max}\leq3/4$.
We first transform formulations of Eq. (\ref{fano1}) and Eq. (\ref{fano2})
into the following form:
\begin{eqnarray}
&&H[P(A=B|a=0)]\nonumber\\
&&\geq \sum_{\lambda_{0},\chi_{0}}{P(\lambda=\lambda_{0},\chi=\chi_{0})H(B|\lambda=\lambda_{0},\chi=\chi_{0})}
\label{fano55}\\
\nonumber\\
&&H[P(A=B\oplus b|a=1)]\nonumber\\
&&\geq \sum_{\lambda_{0},\chi_{0}}{P(\lambda=\lambda_{0},\chi=\chi_{0})H(B\oplus b|\lambda=\lambda_{0},\chi=\chi_{0})},\nonumber\\
\label{fano66}
\end{eqnarray}
where $P(\lambda=\lambda_{0},\chi=\chi_{0})$ is the probability of $\lambda=\lambda_{0},\chi=\chi_{0}$, and $H(B|\lambda=\lambda_{0},\chi=\chi_{0})$
and $H(B\oplus b|\lambda=\lambda_{0},\chi=\chi_{0})$ are conditional entropy about $B$ and $B\oplus b$ respectively
given $\lambda=\lambda_{0},\chi=\chi_{0}$.

Consider the following bound conditions
\begin{eqnarray}
&&H[P(A=B|a=0)]\geq H(B|\lambda=\lambda_{0},\chi=\chi_{0})
\label{fano5}\\
&&H[P(A=B\oplus b|a=1)]\geq H(B\oplus b|\lambda=\lambda_{0},\chi=\chi_{0})\nonumber\\
\label{fano6}
\end{eqnarray}
 We can see that in the condition of $I(b;\lambda,\chi)=0$ if for any given $\lambda_{0},\chi_{0}$ bound conditions of Eq.(\ref{fano5}) and Eq.(\ref{fano6})
make $\beta_{max}\leq3/4$ then bound conditions of Eq.(\ref{fano55}) and Eq.(\ref{fano66}) also demand $\beta_{max}$ less than or equal to $3/4$.

Now we prove that the bound conditions Eq.(\ref{fano5}) and Eq.(\ref{fano6})
make $\beta_{max}\leq3/4$ on the condition of $I(b;\lambda,\chi)=0$. Since $\sum_{\lambda_{0},\chi_{0}}{P(\lambda=\lambda_{0},\chi=\chi_{0})H(b|\lambda=\lambda_{0},\chi=\chi_{0})}
\leq\sum_{\lambda_{0},\chi_{0}}{P(\lambda=\lambda_{0},\chi=\chi_{0})}=1$ and $H(b)=1$, $I(b;\lambda,\chi)=0$ means that for
any $\lambda_{0}, \chi_{0}$ we always have $H(b|\lambda=\lambda_{0},\chi=\chi_{0})=1$. This also means that $P(b=0|\lambda=\lambda_{0},\chi=\chi_{0})=P(b=1|\lambda=\lambda_{0},\chi=\chi_{0})=1/2$ for
any $\lambda_{0}, \chi_{0}$ on the condition of $I(b;\lambda,\chi)=0$. So $H(B;\lambda=\lambda_{0},\chi=\chi_{0})$ can
be written as
\begin{eqnarray}
&&H(B;\lambda=\lambda_{0},\chi=\chi_{0})\nonumber\\
&=&-\sum_{B}{P(B|\lambda=\lambda_{0},\chi=\chi_{0})\log{P(B|\lambda=\lambda_{0},\chi=\chi_{0})}},\nonumber\\
\end{eqnarray}
where
\begin{eqnarray}
&&P(B|\lambda=\lambda_{0},\chi=\chi_{0})=\sum_{b}{P(B,b|\lambda=\lambda_{0},\chi=\chi_{0})}\nonumber\\
&=&\sum_{b}{P(B|b,\lambda=\lambda_{0},\chi=\chi_{0}) P(b|\lambda=\lambda_{0},\chi=\chi_{0})}\nonumber\\
&=&\frac{1}{2}P(B|b=0,\lambda=\lambda_{0},\chi=\chi_{0})\nonumber\\
&&+\frac{1}{2}P(B|b=1,\lambda=\lambda_{0},\chi=\chi_{0})
\label{x1}
\end{eqnarray}
Using the above method we can also calculate $H(B\oplus b;\lambda=\lambda_{0},\chi=\chi_{0})$. If we define
$P(B=0|b=0,\lambda=\lambda_{0},\chi=\chi_{0})=P_{1}$ and $P(B=0|b=1,\lambda=\lambda_{0},\chi=\chi_{0})=P_{2}$
then $P(B=1|b=0,\lambda=\lambda_{0},\chi=\chi_{0})=1-P_{1}$ and $P(B=1|b=1,\lambda=\lambda_{0},\chi=\chi_{0})=1-P_{2}$,
$H(B;\lambda=\lambda_{0},\chi=\chi_{0})$ and $H(B\oplus b;\lambda=\lambda_{0},\chi=\chi_{0})$ can be expressed in terms
of $P_{1},P_{2}$ as follows:
\begin{eqnarray}
&&H(B;\lambda=\lambda_{0},\chi=\chi_{0})=-\frac{P_{1}+P_{2}}{2}\log{\frac{P_{1}+P_{2}}{2}}\nonumber\\
&&-(1-\frac{P_{1}+P_{2}}{2})\log{(1-\frac{P_{1}+P_{2}}{2})},
\label{i1}
\end{eqnarray}
\begin{eqnarray}
&&H(B\oplus b;\lambda=\lambda_{0},\chi=\chi_{0})\nonumber\\
&&=-\frac{1+P_{1}-P_{2}}{2}\log{\frac{1+P_{1}-P_{2}}{2}}\nonumber\\
&&-\frac{1+P_{2}-P_{1}}{2}\log{\frac{1+P_{2}-P_{1}}{2}}
\label{i2}
\end{eqnarray}
\begin{figure}[t]
\includegraphics[width=0.8\columnwidth,
height=0.6\columnwidth]{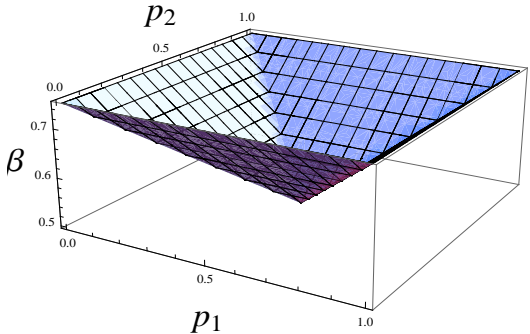} \caption{(Color online). Here axes label $P_{1}$ and $P_{2}$ represent $P(B=0|b=0,\lambda=\lambda_{0},\chi=\chi_{0})$ and
$P(B=0|b=1,\lambda=\lambda_{0},\chi=\chi_{0})$ respectively, for any given $\lambda_{0},\chi_{0}$. We can find $\beta_{max}$ always less than or equal to $3/4$.
$\beta_{max}$ take its minimum $1/2$ in the case of $P_{1}=P_{2}=1/2$ which corresponding
$I(B;\lambda=\lambda_{0},\chi=\chi_{0})=I(B\oplus b;\lambda=\lambda_{0},\chi=\chi_{0})=0$.}
\label{fig1}
\end{figure}
Substitute Eq.(\ref{i1}) and Eq.(\ref{i2}) into Eq.(\ref{fano5}) and Eq.(\ref{fano6}) respectively, we get bound conditions of $P(A=B|a=0)$
and $P(A=B\oplus b|a=1)$ which are both functions of $P_{1}$ and $P_{2}$. We depict $\beta_{max}$ in Fig.(\ref{fig1}). We can find that for any values of
$P_{1}$ and $P_{2}$, the $\beta$ is always less than or equal to $3/4$. It means that for any given $\lambda_{0}$ and $\chi_{0}$ the bound conditions Eq.(\ref{fano5}) and Eq.(\ref{fano6}) make $\beta_{max}\leq3/4$ on the condition of $I(b;\lambda,\chi)=0$. So we finally come to conclusion that $\beta_{max}\leq3/4$ on the condition of $I(b;\lambda,\chi)=0$. 
It's easy to find that the above proof is also valid for the biased case,
the mutual information about distant measurement settings is still required in that case.

In the above we have proven the fact in terms of mutual information:
if Alice and Bob have freedom of measurement choice, in order to violate Bell's inequality nonlocal information must contain both information about distant outcome and measurement setting. The nonlocal information about distant outcome can be obtained directly from hidden variable $\lambda$, but the information about distant measurement
setting must be transmitted by message $\chi$, this fact also can be proved by mutual information. We define the transmitted information about distant setting and
outcome to be $\Delta(b;\chi)=I(b;\lambda,\chi)-I(b;\lambda)$ and $\Delta(B;\chi)=I(B;\lambda,\chi)-I(B;\lambda)$ respectively. Since Bob's measurement choice is free
the mutual information $I(b;\lambda)=0$, this means $\Delta(b;\chi)=0$ will induce $I(b;\lambda,\chi)=0$. We have proven that if $I(b;\lambda,\chi)=0$ then it's
impossible to violate Bell's inequalities, so the information about $b$ must be transmitted. On the other hand, $I(B;\lambda)$ is not usually equal to zero,
so $\Delta(B;\chi)=0$ is allowable. It's easy to find that the asymmetry between $\Delta(b;\chi)$ and $\Delta(B;\chi)$ originates from the freedom
of measurement choice.

Now we discuss the case with biased marginal probabilities.
We suppose $H(B)=h_{1}$ and $H(B\oplus b)=h_{2}$, the bound conditions
of Eq. (\ref{fano3}) and Eq. (\ref{fano4}) transform to
\begin{eqnarray}
&&H[P(A=B|a=0)]\geq h_{1}-I(B;\lambda,\chi);
\label{b1}\\
\nonumber\\
&&H[P(A=B\oplus b|a=1)]\geq h_{2}-I(B\oplus b;\lambda,\chi).
\label{b2}
\end{eqnarray}
In the case with biased marginal probabilities one has $h_{1}<1$, so in Eq. (\ref{b1})
if $I(B;\lambda,\chi)=0$ one can still get $P(A=B|a=0)>\frac{1}{2}$, this means it's still possible to achieve a violation of Bell's inequality.
So the mutual information about distant outcome is no longer necessary.

We can imagine that there are three people: Alice, Bob, and Referee, they will do a Bell experiment, the intention of
this experiment is that the Referee wants to test whether there exists a stronger-than-classical correlation between
Alice and Bob. The design of this experiment is: Referee generates two unbiased random variables $a,b\in\{0,1\}$
and send them to Alice and Bob respectively, and Alice and Bob send their outputs $A,B\in\{0,1\}$ to the Referee.
After many trials, the Referee collects all the data, and if the data results in violation of any one of the
following eight Bell inequalities the Referee will confirm that there exists a stronger-than-classical correlation between
Alice and Bob \cite{1}.
\begin{eqnarray}
\frac{1}{4}\sum_{a,b=0}^{1}{p(A\oplus B= ab\oplus\alpha a\oplus\beta b\oplus\gamma|a,b)}
\leq\frac{3}{4},
\label{new bell ineq}
\end{eqnarray}
where $\alpha,\beta,\gamma \in\{0,1\}$. The choice of $\alpha=0,\beta=0,\gamma=0$ corresponding to the standard CHSH inequality
and is widely used. Now we prove that even $I(B;\lambda,\chi)=0$ it's still
possible to achieve a violation of Bell's inequality as long as Alice has enough information
about $B\oplus b$, in the case of biased marginal probabilities. Alice's strategy is: when she receives $a=0$ she always outputs $A=1$ or $A=0$.
Without losing generality we can assume that Alice always outputs $A=1$ when she receives $a=0$, apart from that she
can always take $A$ equal to $B\oplus b$ when she receives $a=1$ since she has full information of $B\oplus b$.
So if the probability of $B=1$ is greater than $1/2$, the Bell inequality of $\alpha=0,\beta=0,\gamma=0$ will be violated;
and if the probability of $B=0$ is greater than $1/2$, the Bell inequality of $\alpha=1,\beta=0,\gamma=1$ will be violated.
The above two Bell inequality just differ by a local transform: $A\rightarrow A\oplus 1$ and $b\rightarrow b\oplus 1$.
This conclusion is a genuine new result obtained by using mutual information.

\begin{figure}[t]
\includegraphics[width=0.8\columnwidth,
height=0.6\columnwidth]{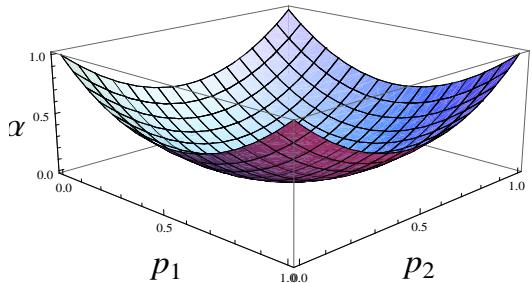} \caption{(Color online). Here axes label $P_{1}$ and $P_{2}$ represent $P(B=0|b=0,\lambda=\lambda_{0},\chi=\chi_{0})$ and
$P(B=0|b=1,\lambda=\lambda_{0},\chi=\chi_{0})$ respectively, for any given $\lambda_{0},\chi_{0}$.
We can find $\alpha_{max}$ always less than or equal to $1$ if $I(b;\lambda,\chi)=0$.}
\label{fig2}
\end{figure}

Finally, we discuss the relation between nonlocal information and the violation of
information causality \cite{pawlowski2}.
In our this model the principle of information causality can be expressed as \cite{pawlowski2}:
\begin{eqnarray}
&&[2P(A=B|a=0)-1]^{2}+[2P(A=B\oplus b|a=1)-1]^{2}\nonumber\\
&&\leq1
\label{inforcau}
\end{eqnarray}
If we define $[2P(A=B|a=0)-1]^{2}+[2P(A=B\oplus b|a=1)-1]^{2}$ as $\alpha$, like $\beta_{max}$ the values of $\alpha_{max}$ also be confined
by the bound conditions Eq.(\ref{fano3}) and Eq.(\ref{fano4}), only when $\alpha_{max}>1$ it's possible to violate information
causality. It's obvious that when $I(B;\lambda,\chi)=0$
the maximum of $P(A=B|a=0)$ is $\frac{1}{2}$, therefore $\alpha_{max}$ can not be greater than $1$.
If $I(b;\lambda,\chi)=0$, $H(B;\lambda=\lambda_{0},\chi=\chi_{0})$ and $H(B\oplus b;\lambda=\lambda_{0},\chi=\chi_{0})$ can
be expressed as Eq. (\ref{i1}) and Eq. (\ref{i2}), we depict the maximum of $\alpha$ under the bound conditions Eq. (\ref{fano5})
and Eq. (\ref{fano6}) in Fig.(\ref{fig2}). We find the maximum of $\alpha$ is always less than $1$. So we can conclude
that the mutual information about both distant outcome and setting are necessary for a violation of information causality.

In summary, we have discussed a nonlocal realism model with an assumption of free and unbiased measurement settings, the nonlocal information of this model
can be expressed as mutual information about the distant outcome and measurement setting, and we find that in this model
the nonlocal information plays a key role. We find that, in the case of unbiased marginal probabilities the mutual information
about both distant outcome and setting are necessary for a violation of Bell's inequality,
and the nonlocal information about distant setting must be
contained in the transmitted message. With respect to the case with biased marginal probabilities, we point out
that the mutual information about distant outcome ceases to be necessary for the violation of Bell's inequality,
while the mutual information about distant measurement settings is still required.
Finally we prove that the mutual information about both distant outcome and setting are necessary for a violation of information causality.


{\it Acknowledgments}
~Y. X. is supported by National Foundation of Natural Science in
China Grant Nos. 10947142 and 11005031.
S.-J. X. is supported by the State Key
Programs for Basic Research of China (Grant Nos.2005CB623605 and
2006CB921803), and by National Foundation of Natural Science in
China Grant Nos. 10874071 and 10704040.




\begin{thebibliography}{100}
\bibitem{bell} J. S. Bell, Physics (Long Island City, N.Y.) {\bf1},
195(1964); J. S. Bell, \emph{Speakable and Unspeakable in Quantum
Mechanics} (Cambridge University Press, Cambridge, England, 1988).
\bibitem{chsh} J. F. Clauser, M. A. Horne, A. Shimony, and R. A.
Holt, Phys. Rev. Lett. {\bf23}, 880(1969); {\bf24}, 549(E)(1970).
\bibitem{toner} B. F. Toner and D. Bacon, Phys. Rev. Lett. {\bf91}, 187904(2003).
\bibitem{pr} S. Popescu and D. Rohrlich, Found. Phys. {\bf24}, 379(1994).
\bibitem{cerf} N. J. Cerf, N. Gisin, S. Massar, and S. Popescu, Phys. Rev. Lett. {\bf94}, 220403 (2005)
\bibitem{shimony} A. Shimony, \emph{Search for a Naturalistic World View}, Vol.2(Cambridge University Press, Cambridge, UK, 1993);
M. P. Seevinck, \emph{Parts and Wholes}, arXiv: 0811.1027
;B. Terhal \emph{et al}, Phys. Rev. Lett. {\bf90}, 157903(2003); G. Brassard, R. Cleve, and A. Tapp, Phys. Rev. Lett. {\bf83}, 1874(1999);
M. Steiner, Phys. Lett. A {\bf270}, 239(2000);
J. A. Csirik, Phys. Rev. A {\bf66}, 014302(2002);
D. Bacon and B. F. Toner, Phys. Rev. Lett. {\bf90}, 157904(2003);
N. J. Cerf, N. Gisin, and S. Massar, Phys. Rev. Lett. {\bf84}, 2521(2000);
S. Massar, D. Bacon, N. Cerf, and R. Cleve, Phys. Rev. A {\bf63}, 052305(2001).
\bibitem{pawlowski1} M. Paw{\l}owski, J. Kofler, T. Paterek, M. Seevinck, and \v{C}. Brukner, arXiv: 0903.5042v2.
\bibitem{pawlowski2} M. Paw{\l}owski, T. Paterek, D. Kaszlikowski, V. Scarani, A. Winter, and M. \.{Z}ukowski, Nature {\bf461}, 1101(2009);
arXiv: 0905.2292v1(2009).

\bibitem{cover} T. M. Cover and J. A. Thomas, \emph{Elements of Information Theory}(China Machine Press, Beijing, China, 2008).

\bibitem{1} J. Barrett, N. Linden, S. Massar, S. Pironio, S. Popescu, and D. Roberts, Phy. Rev. A {\bf71}, 022101(2005)



\end{thebibliography}
\end{document}